\documentclass[11pt,a4paper]{article}

\usepackage{jheppub}

\usepackage{graphicx, epsfig} 
\usepackage{amsmath,amssymb,amsfonts,dsfont,mathrsfs,amsthm,mathtools, cancel}
\usepackage{bm} 
\usepackage{color}
\usepackage[usenames]{xcolor}
\usepackage{hyperref}
\usepackage{physics}
\hypersetup{colorlinks=true,urlcolor=blue,linkcolor=magenta,citecolor=blue,filecolor=blue}
\usepackage[normalem]{ulem}
\usepackage{array}
\usepackage{hyperref}
\usepackage{booktabs}
\usepackage{blindtext}
\usepackage{lipsum}  
\usepackage{slashed}
\usepackage{amsthm}
\usepackage{orcidlink}

\allowdisplaybreaks

\usepackage{float}
\usepackage[scr=rsfs]{mathalfa}
\usepackage{bigints}

\newcommand{\Lag}{\mathscr{L}}

\newtheorem{theorem}{Theorem}

\begin{document}

\title{Merons beyond Yang-Mills theory}

\author[a]{Borja Diez\,\orcidlink{0009-0004-3805-4036},}
\author[b]{Luis Guajardo\,\orcidlink{0000-0001-7591-7233}}

\affiliation[a]{Departamento de F\'{\i}sica, Cinvestav, Av.~IPN 2508, 07360, CDMX, M\'exico
\vspace{0.1cm}}
\affiliation[b]{Instituto de Matem\'atica, F\'isica y Estad\'istica, Facultad de Ingenier\'ia y Negocios,
Universidad de las Am\'ericas, Sede Concepci\'on,
Avenida Jorge Alessandri Rodríguez 1160, 4090940, Chile}

\vspace{0.3cm}
\emailAdd{borja.diez@cinvestav.mx}
\emailAdd{luis.guajardo.r@gmail.com}

\abstract{
Within the wide variety of topological solitons supported by Yang--Mills theory, merons occupy a particularly distinguished role. Despite their simplicity, they represent genuinely non-Abelian configurations that can be regarded as the fundamental building blocks of instantons, and they provide a qualitatively accurate picture of confinement. In this work, we show that such configurations are, in fact, supported by a broad class of non-Abelian gauge theories beyond Yang--Mills, provided that suitable physical conditions are satisfied. Taking into account their gravitational backreaction, we further demonstrate that both  black holes and Euclidean wormholes sourced by merons admit natural extensions within this generalized framework, which regularizes the singular behavior they exhibit in constant--curvature backgrounds. As a byproduct, we construct a regular black hole solution supported by genuinely non-Abelian gauge fields, based on a non-Abelian generalization of the Ay\'on--Beato--Garc\'ia nonlinear electrodynamics. As a consequence, physical effects intrinsic to merons are likewise expected to be inherited by these classes of theories beyond the Yang-Mills framework. A notable example is the spin from isospin effect, whereby bosonic excitations charged under the gauge group can effectively behave as fermionic degrees of freedom.
}

\maketitle

\section{Introduction\label{sec:intro}}
Non-Abelian gauge theories play a fundamental role in modern theoretical physics, ranging from the Standard Model of particle physics to a wide class of extensions inspired by quantum gravity and string theory. In particular, Yang--Mills theory admits a rich spectrum of non-perturbative configurations---such as monopoles, instantons and merons---whose existence is intimately tied to the topology of the gauge group and the nonlinear structure of the field equations. These configurations encode essential physical information, from tunneling processes in quantum field theory to the structure of strongly coupled phases and confinement.

Among the simplest genuinely non-Abelian configurations are the so-called merons, originally introduced by de Alfaro, Fubini and Furlan~\cite{deAlfaro:1976qet} as solutions of the Yang--Mills equations in Euclidean flat space. These configurations are proportional to a pure gauge potential; however, due to the non-Abelian nature of the theory, the associated field strength does not vanish. This striking feature has no analogue in Abelian gauge theories such as Maxwell electrodynamics. Unlike instantons, merons typically exhibit singular cores, but they carry fractional topological charge and play an important role in semiclassical analyses of Yang--Mills theory, particularly in connection with confinement mechanisms~\cite{Callan:1977qs,Callan:1977gz,Callan:1978bm,Polyakov:1976fu,Actor:1979in,Negele:1998ev,Steele:2000xk}. Their relative simplicity makes them especially useful for constructing analytic solutions in situations where solving the full Yang--Mills equations would otherwise be intractable.

The interplay between non-Abelian gauge fields and gravity provides a natural framework in which these configurations can acquire further physical significance. In particular, the Einstein--Yang--Mills system admits black hole and soliton solutions that violate the traditional no-hair conjecture~\cite{Bizon:1990sr}. While many such solutions have been obtained numerically~\cite{Smoller:1997qr,Winstanley:2008ac,Volkov:1998cc,Shepherd:2015dse,Shepherd:2016ily}, analytic examples are comparatively scarce. A remarkable example is the self-gravitating meron solution constructed in Ref.~\cite{Canfora:2012ap}, where the Yang--Mills field retains its meronic form while the spacetime geometry corresponds to a Reissner--Nordstr\"om black hole with an effective charge of purely non-Abelian origin. Despite the spherical symmetry of the energy-momentum tensor, the gauge field remains intrinsically non-Abelian. 

This construction has been extended in several directions, including higher-rank gauge groups such as $SU(N)$~\cite{Canfora:2022nso} (see also Ref.~\cite{Aviles:2026fsf}), higher dimensions~\cite{Canfora:2018ppu,Flores-Alfonso:2020ayc}, and more general gravitational theories. In particular, analytic examples of self-gravitating anisotropic merons have been recently obtained in Refs.~\cite{Canfora:2023bug,Corral:2024xfv}, where the gauge fields backreact on the geometry in a nontrivial way, producing deformed backgrounds whose structure is dictated by the gauge potential itself. Furthermore, meronic configurations have also been shown to exist in Conformal Gravity, leading to extensions of the Riegert black hole~\cite{2412.08734}. These results highlight the robustness of merons as sources of nontrivial geometries and their relevance in the study of gravitating solitons in differents setups.

Another class of solutions arising in the Einstein--Yang--Mills system is provided by Euclidean wormholes supported by non-Abelian gauge fields. Euclidean wormholes have long been regarded as important semiclassical configurations in quantum gravity, as they can induce nontrivial effects such as the renormalization of coupling constants and the generation of effective interactions~\cite{Coleman:1988tj,Giddings:1988cx}. In particular, they have been proposed as a possible mechanism to address the cosmological constant problem through the Coleman mechanism~\cite{Coleman:1988tj,Hawking:1987mz}. These configurations can be interpreted as instantons mediating topology change, connecting otherwise disconnected asymptotic regions of spacetime~\cite{Giddings:1987cg}. At distances much larger than the wormhole throat, their effects can be encoded in local operator insertions in the effective theory, leading to nonperturbative modifications of the infrared dynamics and inducing interactions among light fields~\cite{Coleman:1989zu,Giddings:1989bq}. These contributions are closely related to the notion of baby universes and may have profound implications for predictivity and the structure of the vacuum in quantum gravity. More recently, Euclidean wormholes have experienced renewed interest in the context of holography and the anti-de Sitter/conformal field theory (AdS/CFT) correspondence~\cite{Maldacena:1997re,Witten:1998qj,Gubser:1998bc}, where they are connected to ensemble averages and factorization puzzles in quantum gravity~\cite{Maldacena:2004rf,Stanford:2020wkf}. In particular, AdS wormholes provide a controlled setting in which these ideas can be explored within a holographic framework. A wide variety of such configurations have been constructed in different settings, including supergravity theories~\cite{Hertog:2017owm,Anabalon:2018rzq,Anabalon:2020loe,Anabalon:2023kcp,Astesiano:2023iql}, models involving scalar fields~\cite{Blazquez-Salcedo:2020nsa,Betzios:2024zhf,Anabalon:2012tu,Cisterna:2021xxq,Barrientos:2022avi,Cisterna:2023uqf}, Skyrmions~\cite{Canfora:2025roy}, and non-Abelian gauge fields~\cite{Hosoya:1989zn,Chew:2021vxh,Canfora:2025gwm,Betzios:2017krj}.

The study of universality in field theories is strongly motivated by the search for structures that remain robust under deformations of the underlying dynamics. In many physical contexts (ranging from effective field theories to quantum gravity and string theory) the action principle is only known perturbatively or is expected to receive higher-order corrections whose precise form is often uncertain. Universal configurations provide a fundamental window into this problem, as they correspond to saddle points that solve not just a particular theory, but a wide class of theories sharing the same basic principles, such as locality and covariance. As a result, these configurations capture features of the theory that are insensitive to microscopic details, thereby offering nonperturbative insights into sectors that would otherwise be difficult to access. In this sense, such geometries are said to be \textit{quantum protected}, making them of considerable theoretical interest. A paradigmatic example of such configurations is provided by $pp$-wave spacetimes in gravitational theories, which are characterized by the vanishing of all scalar curvature invariants. This property ensures that they remain exact solutions of a wide class of gravitational theories, even in the presence of higher-curvature corrections~\cite{Amati:1988sa,Horowitz:1989bv,Coley:2008th,Coley:2002gz,Gurses:2014soa}. Recently, this notion has been further extended to the concept of almost universal spacetimes. In this context, based on $pp$-waves constructed over base manifolds of constant curvature, whose backgrounds are non-Einsteinian, it was shown that, for generic higher-curvature theories, the field equations reduce to two algebraic relations together with a linear differential equation governing the Kerr-Schild profile~\cite{Gurses:2026cat}. Another notable example of universality is given by Skyrmions (topological solitons of the Skyrme model, classified by an integer winding number). In particular, self-gravitating Skyrmion solutions, originally constructed for $SU(2)$ in Ref.~\cite{Ayon-Beato:2015eca} and later extended to $SU(3)$ in Ref.~\cite{Ayon-Beato:2019tvu}, have recently been shown to admit a systematic embedding into larger symmetry groups, such as $SU(N)$. This construction preserves their topological character while extending their validity even in the presence of higher-order corrections in the 't Hooft expansion, and allows for arbitrary values of the topological charge~\cite{Vera:2025qqz}. The chiral soliton lattice exhibit a similar universal behavior~\cite{Canfora:2025aau}. More generally, such universal configurations often exhibit enhanced geometric or algebraic properties that underlie their existence, making them valuable probes in both the classical and quantum regimes of field theories.

On the other hand, modern theoretical physics encompasses a wide range of extensions of gauge dynamics, including nonlinear electrodynamics, higher-order gauge interactions, and effective field theories motivated by quantum corrections~\cite{Dowker:1975tf,Duff:1975ue,Dunne:2004nc} or string theory~\cite{Gross:1986mw,Gross:1986iv,Metsaev:1987zx,Tseytlin:1997csa}(see also Ref.~\cite{Burgess:2007pt} and references therein). Within these frameworks, it is natural to ask whether meronic configurations continue to be supported and, if so, how their properties are modified once gravitational backreaction is taken into account. The aim of this work is to address these questions by studying merons in a broad class of non-Abelian gauge theories beyond Yang--Mills, constructed from the quadratic gauge invariants $X$ and $Y$ [see Eq.~\eqref{X-Y} below]. In particular, we show that merons persist in this generalized setting and give rise to new families of self-gravitating solutions, including meronic black holes and Euclidean wormholes, which help regularize the singular core that these solutions exhibit in constant curvature spacetimes. These results provide further evidence that merons can be extended to a broad class of non-Abelian gauge theories beyond Yang-Mills, with deep connections to both gravitational physics and nonperturbative gauge dynamics.

The remainder of this paper is organized as follows. In Sec.~\ref{sec:Non-Abelian gauge theories}, we present the dynamics of the class of non-Abelian gauge theories under consideration and fix our conventions. In Sec.~\ref{sec:merons on constant curvature spaces}, we analyze the conditions under which merons are supported by theories beyond Yang--Mills on Euclidean spaces of constant curvature. Then, in Sec.~\ref{sec:Self-gravitating merons}, we include the gravitational backreaction of these configurations and show that both meronic black holes and meronic Euclidean wormholes are supported within this broader class of theories. Finally, we conclude with a discussion of our results and possible extensions of this work in Sec.~\ref{sec:Discussion}.

\section{Non-Abelian gauge theories}\label{sec:Non-Abelian gauge theories}
In this work we consider non-Abelian gauge theories for the simplest non-Abelian gauge group, namely $SU(2)$.  To this end, we parametrize the generators of the group as $t_a=-\frac{i}{2}\tau_a$, where $\tau_a$ denote the Pauli matrices, which satisfy 
\begin{equation}
    \tau_a\tau_b=\delta_{ab}\mathbb{I}+i\epsilon_{abc}\tau^c\,.
\end{equation}
From this relation it follows that the generators obey
\begin{equation}\label{comm-trace}
    [t_a,t_b]=\epsilon_{abc}t^c\,,\qquad 
    \Tr(t_at_b)=-\frac{1}{2}\delta_{ab}\,,
\end{equation}
where $\delta_{ab}$ and $\epsilon_{abc}$ are the two invariant tensors of $SU(2)$. From now on, Latin indices will denote $SU(2)$ indices, while Greek indices will refer to spacetime indices. The non-Abelian gauge potential in the adjoint representation of $SU(2)$ is defined as $A_\mu=A_\mu^{a}t_a$ with associated non-Abelian field strength
\begin{equation}
    F_{\mu\nu}=\partial_\mu A_\nu-\partial_\nu A_\mu+[A_\mu,A_\nu]\,,
\end{equation}
which satisfies the Bianchi indentity
\begin{equation}
    \nabla_\mu \tilde{F}^{\mu\nu}+[A_\mu,\tilde{F}^{\mu\nu}]=0\,,
\end{equation}
where $\tilde{F}_{\mu\nu}=\frac{1}{2}\epsilon_{\mu\nu\lambda\rho}F^{\lambda\rho}$ is the dual field strength and $\epsilon_{\mu\nu\lambda\rho}$ denotes the Levi-Civita tensor.

We shall consider arbitrary gauge theories constructed from the two quadratic gauge invariants built out of the non-Abelian field strength,
\begin{equation}\label{X-Y}
    X=\frac{1}{2}\Tr(F_{\mu\nu}F^{\mu\nu})\qquad \text{and}
    \qquad 
    Y=\frac{1}{2}\Tr(\tilde{F}_{\mu\nu}F^{\mu\nu})\,,
\end{equation}
which are even and odd under parity transformations, respectively.

The dynamics of the gauge fields is governed by the action principle
\begin{equation}\label{action}
    I_{\rm nA}=\int_\mathcal{M}\dd^4x\sqrt{|g|}\,\Lag(X,Y)\,,
\end{equation}
where $\Lag(X,Y)$ is a nonlinear Lagrangian density depending only on the invariants $X$ and $Y$, and therefore on first derivatives of the gauge potential.

Stationary variations of the action~\eqref{action} with respect to the gauge connection $A_\mu$ yield the equations of motion
\begin{equation}\label{eom-A}
    \nabla_\mu P^{\mu\nu}+[A_\mu,P^{\mu\nu}]=0\,,
\end{equation}
where $P_{\mu\nu}$ are the non-Abelian constitutive relations,
\begin{equation}\label{P}
    P_{\mu\nu}=\pdv{\Lag}{X}F_{\mu\nu}+\pdv{\Lag}{Y}\tilde{F}_{\mu\nu}
    \equiv L_X F_{\mu\nu}+L_Y\tilde{F}_{\mu\nu}\,.
\end{equation}
The action in Eq.~\eqref{action} includes Yang--Mills when $\Lag=X$, as well as the non-Abelian extensions of Euler-Heinsenberg~~\cite{Dowker:1975tf,Duff:1975ue,Dunne:2004nc}, ModMax~\cite{Cirilo-Lombardo:2023poc,Canfora:2025gwm}, Born-Infeld~\cite{Tseytlin:1997csa}, among others.

It is worth emphasizing that the gauge group $SU(2)$ admits a richer set of gauge-invariant quantities than those considered here. Nevertheless, throughout this work we restrict our attention to the class of non-Abelian theories constructed solely from the invariants appearing in Eq.~\eqref{X-Y}. These correspond to the simplest non-Abelian generalizations of the only two independent gauge invariants that can be built in the Abelian case.

\section{Merons on constant curvature spaces}\label{sec:merons on constant curvature spaces}
One of the most remarkable properties of merons is that they correspond to genuinely non-Abelian solitons of Yang--Mills theory on Euclidean flat space~\cite{deAlfaro:1976qet}. In this work we show that, in fact, such configurations are admitted by a broad class of non-Abelian gauge theories satisfying reasonable physical assumptions.

To this end, we consider Euclidean spaces of constant curvature $k=\{-1,0,1\}$, corresponding respectively to $\mathbb{H}^4$, $\mathbb{R}^4$, and $\mathbb{S}^4$. The line element of these spaces can be parametrized as
\begin{equation}\label{ds-k}
    \dd s^2=\dd r^2+\frac{\sin^2(\sqrt{k}r )}{4k}(\sigma_1^2+\sigma_2^2+\sigma_3^2)\,,
\end{equation}
where $r$ is a radial coordinate whose range is $r\in(0,\pi]$ for $k=1$ and $r\in[0,\infty)$ otherwise. The $\sigma_a$ are the left-invariant one-forms (LIF) of $SU(2)$ defined by
\begin{subequations}
	\begin{align}
  \sigma_1&=\sin\psi\dd\vartheta - \sin\vartheta\cos\psi\dd\varphi\,,\\
    \sigma_2&=\cos\psi\dd\vartheta + \sin\vartheta\sin\psi\dd\varphi\,,\\
    \sigma_3&=\dd\psi + \cos\vartheta\dd\varphi\,,
\end{align}
\end{subequations}
which satisfy the Maurer-Cartan relation
$\dd\sigma_a+\frac{1}{2}\epsilon_{abc}\sigma^b\wedge\sigma^c=0$. The coordinates $(\vartheta,\varphi,\psi)$ are Euler angles with ranges $\vartheta\in[0,\pi)$, $\varphi\in[0,2\pi)$, and $\psi\in[0,4\pi]$.

For the non-Abelian gauge field we adopt an ansatz aligned with the LIF of $SU(2)$, namely,
\begin{equation}\label{A-meron}
	A=A_\mu^{a}\dd x^\mu t_a=\lambda\sigma^{a}t_a\,,
\end{equation}
where $\lambda$ is a constant to be determined by the equations of motion. The associated field strength is given by
\begin{equation}
\label{F}
    F=\frac{1}{2}F_{\mu\nu}\dd x^\mu\wedge \dd x^\nu=\frac{\lambda(\lambda-1)}{2}\epsilon^{a}_{~bc}\sigma^b\wedge \sigma^c t_a\,.
\end{equation}
As usual, $\lambda=1$ corresponds to a pure gauge configuration, since the corresponding field strength vanishes identically. Thus, in what follows we will be interested in configurations with $\lambda\neq\{0,1\}$.
It is worth emphasizing that in Abelian theories, such as Maxwell electrodynamics, gauge fields of the form~\eqref{A-meron} are pure gauge for any value of $\lambda$. For this reason, despite their simplicity, merons constitute genuinely non-Abelian configurations.

A direct computation of the gauge invariants defined in Eq.~\eqref{X-Y} shows that, for the ansatz~\eqref{A-meron}, they take the form
\begin{equation}
\label{eq:meron}
    X=-\frac{24 k^2\lambda^2(\lambda-1)^2}{\sin(\sqrt{k}r)^4}\,\qquad \text{and}\qquad Y=0\,.
\end{equation}
That is, on shell, the Lagrangian $\Lag$ becomes a function that depends solely on $X$. Let us note that the flat-space limit, i.e., $k\to 0$, of this invariant is smooth, which allows us to treat all three geometries on the same footing.

The non-Abelian constitutive relations~\eqref{P} acquire both electric and magnetic components, namely,
\begin{equation}
	P=\frac{1}{2}P_{\mu\nu}\dd x^\mu \wedge\dd x^\nu = \lambda(\lambda-1)\left[\frac{2\sqrt{k}L_Y}{\sin(\sqrt{k}r)}\dd r\wedge\sigma^{a}+\frac{L_X}{2}\epsilon^{a}_{~bc}\sigma^b\wedge\sigma^c\right]t_a\,.
\end{equation}
Substituting these expressions into the field equations~\eqref{eom-A}, we obtain
\begin{equation}
\label{eq:reducedfieldeq}
    \frac{2k\lambda(\lambda-1)}{\sin(\sqrt{k}r)^2}\left[L_Y'(r)-\frac{2\sqrt{k}(2\lambda-1)}{\sin(\sqrt{k}r)}L_X(r)\right]=0\,,
\end{equation}
where the prime denotes differentiation with respect to the radial coordinate $r$. The latter is solved by
\begin{equation}\label{lsol-dLyrsol}
    \lambda=\frac{1}{2}\qquad \text{and}\qquad \dv{L_Y}{r}=0\,.
\end{equation}

The above derivation relies on the existence of homogeneous $SU(2)$ orbits, and the alignment of the gauge connection with the corresponding LIF. Consequently, it can be straightforwardly extended to any four-dimensional manifold admitting a foliation by $\mathbb{S}^3 \simeq SU(2)$ orbits. This observation leads to the following theorem:

\begin{theorem}
\label{theorem}
Let $\mathcal{M}$ be a four--dimensional manifold foliated by homogeneous $\mathbb{S}^3$ orbits, described by the class of metrics
$$ds^2 = \dfrac{dr^2}{F(r)}+ \dfrac{G(r)}{4}(\sigma_1^2 + \sigma_2^2 + \sigma_3^2)\,,$$ where $\sigma_a$, with $a=1,2,3$, are the LIF of $SU(2)$. Consider $\Lag(X,Y)$ as the Lagrangian density for an $SU(2)$ gauge theory, where $X$ and $Y$ are the quadratic gauge invariants
$$X=\frac{1}{2}\Tr(F_{\mu\nu}F^{\mu\nu})\qquad \text{and}
    \qquad 
    Y=\frac{1}{2}\Tr(\tilde{F}_{\mu\nu}F^{\mu\nu})\,. $$ Then, the gauge connection
\begin{equation}\label{A-theorem}
    A=\frac{1}{2}\sigma_{a}t^a\,,
\end{equation}
satisfies identically the field equations~\eqref{eom-A}, provided that 
\begin{equation}\label{condition}
    \left. \frac{\partial \Lag}{\partial Y}\right|_{Y=0} = {\rm{cst.}}
\end{equation} 
\end{theorem}

This special form of the non-Abelian gauge field is precisely the one that appears in standard Yang--Mills theory~\cite{deAlfaro:1976qet}. However, we see that these configurations are also admitted by a broad class of non-Abelian theories beyond Yang--Mills, provided that the condition in Eq.~\eqref{condition} is satisfied. From the perspective of effective field theories, this requirement is quite natural, since corrections to the Yang--Mills action typically appear as polynomial functions of the gauge invariants, as in the case of non-Abelian Euler--Heisenberg theories~\cite{Dowker:1975tf,Duff:1975ue}, for example (see Ref.~\cite{Dunne:2004nc} for a review). Another criterion to single out theories satisfying Eq.~\eqref{condition} is to impose parity invariance. This requirement is well motivated, as it is consistent with the symmetry properties of Yang--Mills and gravitational interactions, and naturally excludes contributions involving parity-odd terms in the action. Examples of such theories include the non-Abelian extensions of ModMax electrodynamics~\cite{Bandos:2020jsw,Cirilo-Lombardo:2023poc,Canfora:2025gwm} and non-Abelian Born--Infeld theory~\cite{Tseytlin:1997csa}, among others. Thus, physically reasonable theories generically satisfy the condition \eqref{condition}, thereby ensuring the existence of merons within a broad class of non-Abelian gauge theories possesing parity invariance.

In Yang--Mills theory, these configurations are singular at the origin, regardless of the topology of the background or the specific form of the theory, since the non-trivial gauge invariant diverges at that point [see Eq.~\eqref{eq:meron}]. Consequently, on constant-curvature spaces, an isolated meron does not contribute to the path integral, as its Euclidean action is divergent. However, this divergence can be cured in certain theories beyond Yang--Mills. A remarkable example is the non-Abelian generalization of the Ay\'on--Beato--Garc\'ia (ABG) nonlinear electrodynamics~\cite{Ayon-Beato:2000mjt}, whose Lagrangian is given by\footnote{According to the conventions employed in this work, the trace of the generators yields an overall minus sign [cf. Eq.~\eqref{comm-trace}], ensuring that the square roots in Eq.~\eqref{non-Abelian ABG} are well defined.}

\begin{equation}\label{non-Abelian ABG}
	\Lag(X)=-\left(\frac{e\sqrt{-2X}}{1+e\sqrt{-2X}}\right)^q\,,
\end{equation}
where $e\geq 0$ is a coupling constant.

The on-shell action can be expressed in closed form for the three constant--curvature backgrounds as
\begin{equation}
	I_{\rm nA}=
	\begin{cases}
	-\dfrac{8\pi^2}{3}\,_2F_1\left(q,2;\dfrac{5}{2};-\dfrac{1}{e\sqrt{3}}\right)\,, & \text{for }k=1 \,,\\ \\
		-\dfrac{3e^2\pi^2}{(q-1)(q-2)}\,, & \text{for }k=0\,,\\ \\
		-\dfrac{\pi^2\Gamma\left(q-\frac{3}{2}\right)}{\Gamma\left(q+\frac{1}{2}\right)} \,_2F_1\left(q,2;q+\dfrac{1}{2};1-\dfrac{1}{e\sqrt{3}}\right)\,, & \text{for }k=-1\,,
	\end{cases}
\end{equation}
where $\,_2F_1(a,b;c;z)$ denotes the Gauss hypergeometric function, which is defined in terms of the series~\cite{gil2007numerical}
\begin{equation}\label{2F1}
	\,_2F_1(a,b;c;z)=\sum_{j=0}^\infty \frac{(a)_j(b)_j}{(c)_j}\frac{z^j}{j!}\,,
\end{equation}
with $(a)_j$ being the Pochhammer symbols, given by
\begin{equation}
	(a)_j=\frac{\Gamma(z+j)}{\Gamma(z)}=z(z+1)\cdots (z+j-1)\,,
\end{equation}
for $j\in \mathbb{Z}_{>0}$, and $(a)_0=1$. The bound on $q$ ensuring that the action converges depends on the background under consideration. One needs $q>\frac{3}{2}$ for $\mathbb{H}^4$, and $q>2$ for flat space. The integral converges for all $q$ in the $\mathbb{S}^4$ case.

The Abelian version of the theory defined in Eq.~\eqref{non-Abelian ABG} has been employed to support the Bardeen regular black hole model~\cite{isz07} as an exact solution of a nonlinear electrodynamic coupled to gravity, for the specific value $q=\frac{5}{2}$. In that case, the solution can be reinterpreted as the gravitational field generated by a nonlinear magnetic monopole~\cite{Ayon-Beato:2000mjt}. Here, we show that in its non-Abelian extension, and by promoting the parameter $q$ to take arbitrary values, one obtains a broader class of non-Abelian gauge theories that regularize the action for merons on constant-curvature backgrounds.

\section{Self-gravitating merons}\label{sec:Self-gravitating merons}
Motivated by the interest in exploring gravitational solitons beyond standard Yang--Mills theory, in this section we seek self-gravitating meronic configurations in Einstein gravity minimally coupled to the non-Abelian theory defined by the action in Eq.~\eqref{action}, that is,
\begin{equation}
\label{eq:action}
	I[g,A]=\frac{1}{2\kappa}\int_\mathcal{M}\dd^4x\sqrt{|g|}\left(R-2\Lambda\right) + I_{\rm nA}\,,
\end{equation}
where $\kappa=8\pi G$ is the gravitational coupling constant, $\Lambda$ is the cosmological constant, $g=\det g_{\mu\nu}$ is the determinant of the metric, and $R=g^{\mu\nu}R^\lambda_{~\mu\lambda\nu}$ is the Ricci scalar. The equations of motion are obtained by performing arbitrary variations with respect to the metric, yielding
\begin{equation}\label{Einstein-eqs}
	G_{\mu\nu}+\Lambda g_{\mu\nu}=\kappa T_{\mu\nu}\,,
\end{equation}
where stress--energy tensor associated with the non-Abelian fields is given by
\begin{equation}\label{Tmunu}
	T_{\mu\nu}= \Lag (X,Y)g_{\mu\nu}-2\Tr(F_{(\mu}^{~\lambda}P_{\nu)\lambda})\,.
\end{equation}
As discussed in the previous section, meronic configurations are singular at the origin when considered on fixed backgrounds of constant curvature. In what follows, we show that once gravitational backreaction is taken into account, the nontrivial topology of the resulting spacetimes can effectively regularize these configurations. In particular, the singular behavior of merons can be hidden behind horizons or resolved by the global structure of the geometry, rendering them physically acceptable configurations that may contribute, for instance, to the gravitational path integral.

\subsection{Meronic black holes}
To construct meronic black hole solutions, we consider a static and spherically symmetric ansatz for the metric,
\begin{equation}
\label{eq:ansatz-regbh-d4}
	\dd s^2=-f(r)\dd t^2+\frac{\dd r^2}{f(r)}+r^2(\dd\vartheta^2+\sin^2\vartheta\dd\varphi^2)\,,
\end{equation}
while for the gauge field we adopt a meronic ansatz of the form~\cite{Canfora:2012ap},
\begin{equation}
	A=\lambda U^{-1}\dd U\,,
\end{equation}
where $U\in SU(2)$ is parametrized as
\begin{equation}
	U=e^{-\varphi t_1}e^{2\vartheta t_2}e^{\varphi t_3}\,.
\end{equation}
Explicitly, the gauge potential reads
\begin{equation}
	A=2\lambda(\sin\varphi\dd\vartheta+\cos\vartheta\cos\varphi\sin\vartheta\dd\varphi)t_1 + 2\lambda (\cos\varphi\dd\vartheta-\cos\vartheta\sin\vartheta\sin\varphi\dd\varphi)t_2 + 2\lambda \sin^2\vartheta\dd\varphi t_3\,.
\end{equation}
A straightforward computation shows that, for this ansatz, the gauge invariant $Y$ defined in Eq.~\eqref{X-Y} vanishes identically, while
\begin{equation}\label{X-meronic-BH}
	X=-\frac{8\lambda^2(\lambda-1)^2}{r^4}\,.
\end{equation}
From this expression it follows that, in order to avoid trivial configurations, one must require $\lambda\neq\{0,1\}$, as in the case discussed in the previous section. Under this condition, the non-Abelian gauge field equations~\eqref{eom-A} are also satisfied provided that Eq.~\eqref{lsol-dLyrsol} holds. On the other hand, the Einstein equations~\eqref{Einstein-eqs} reduce to a system of two ordinary differential equations, given by
\begin{subequations}
	\label{eq:rel f-Lag}\begin{align}
		\frac{f'(r)}{r}+\frac{f(r)-1}{r^2}+\Lambda-\kappa \Lag &=0\,,\\
		\frac{f''(r)}{2}+\frac{f'(r)}{r}+\Lambda-\frac{\kappa(r^4\Lag + L_X)}{r^4}&=0\,.
	\end{align}
\end{subequations}
Formally, this system is solved by
\begin{equation}
	f(r)=1-\frac{2mG}{r}-\frac{\Lambda r^2}{3}+\frac{\kappa}{r}\int^r \rho^2\Lag(X) \dd \rho\,,
\end{equation}
where $m$ is an integration tipically related with the mass of the solution. In addition, the expression above solves the Einstein equations provided that
\begin{equation}\label{condition-meronic-BH}
	\dv{r}\Lag(X)-\frac{2L_X}{r^5}=0\,.
\end{equation}
However, using the on-shell expression for the invariant $X$ given in Eq.~\eqref{X-meronic-BH}, it is straightforward to verify that Eq.~\eqref{condition-meronic-BH} is automatically satisfied. In other words, all non-Abelian gauge theories constructed from the quadratic gauge invariants in Eq.~\eqref{X-Y} admit meronic black hole solutions, as long they satisfy Eq.~\eqref{lsol-dLyrsol}.

As a consequence, physical effects characteristic of merons are expected to be inherited by those theories satisfying the condition in Eq.~\eqref{lsol-dLyrsol}. A notable example is the spin from isospin effect~\cite{Jackiw:1976xx,Hasenfratz:1976gr}, according to which excitations of bosonic fields charged under the $SU(2)$ gauge group can effectively behave as fermionic degrees of freedom.

Although the gauge invariant $X$ is singular at $r=0$ (a feature already present for merons in constant--curvature spaces) this singularity is hidden behind the event horizon, provided that the latter exists. This suggests that such configurations may, in principle, be physically relevant and observable in isolation. Moreover, these configurations cannot be transformed into an Abelian sector of the theory by any well-defined global gauge transformation, and therefore represent genuinely non-Abelian configurations (see Ref.~\cite{Canfora:2012ap} for further details.)

The existence of a black hole horizon can be ensured by appropriately choosing the Lagrangian $\Lag(X)$, which controls the radial profile of the metric function $f(r)$, thereby guaranteeing that it possesses at least one real positive root.

\subsubsection{ABG meronic black hole}
As an illustrative example, let us consider the non-Abelian extension of the ABG theory defined in Eq.~\eqref{non-Abelian ABG}. In this case, the metric function takes the form
\begin{equation}
	f(r)=1-\frac{2mG}{r} - \frac{r^2}{3}\left[\Lambda + \,_2F_1\left(\frac{3}{2},q;\frac{5}{2},-\frac{r^2}{e}\right)\kappa\right]\,.
\end{equation}

In the limit $e\to 0$, the Schwarzschild-(A)dS solution is recovered. For $q=5/2$, the geometry is effectively reduced to the Bardeen metric \cite{isz07}. The asymptotic behavior of the metric function is given by
\begin{equation}\label{asympt-f-BH}
	f(r)\sim 1-\frac{2mG}{r}-\frac{\Lambda r^2}{3}-\frac{\kappa e^{3/2}\sqrt{\pi}}{4r}\frac{\Gamma(q-\frac{3}{2})}{\Gamma(q)}+\mathcal{O}\left(r^{-2(q-1)}\right)\,,
\end{equation}
where we recall that $q>\frac{3}{2}$. From Eq.~\eqref{asympt-f-BH} it is clear that the ADM  mass (for $\Lambda=0$) is shifted due to the presence of the non-Abelian fields, that is,
\begin{equation}
	M_{\rm ADM}=m+\frac{(\pi e)^{3/2}\Gamma(q-\frac{3}{2})}{\Gamma(q)}\,.
\end{equation}
Note that the stress-energy tensor has the following form
\begin{align}
    T_{\mu}^{\nu} = {\rm{diag}} (\Lag, \Lag, \Lag - 2XL_X, \Lag - 2XL_X)\,,
\end{align}
with
\begin{align*}
    \Lag = -\left(\dfrac{e}{r^2+e}\right)^q\,,\quad L_X = \dfrac{r^6qe^q}{(r^2+e)^{q+1}}\,.
\end{align*}
If one ask for finiteness of the algebraic curvature invariants at $r=0$, the metric function must behave as $f(r)\sim 1 +  O(r^2)$. As elaborated in \cite{Dymnikova:1992ux, Zaslavskii:2010qz} (see also \cite{Dymnikova:2001fb}), coupling matter while keeping this behavior is possible when $T_{\mu}^{\nu} \sim \alpha \delta_{\mu}^{\nu}$. This is precisely the case for the non-Abelian ABG Lagrangian, which saturates near the core as $\Lag \to -1$ and $2XL_X \to 0$, so that $T_{\mu}^{\nu}\to -\delta_{\mu}^{\nu}$. Hence, the stress-energy tensor remains finite at $r=0$; consequently, despite $X$ is singular, when $m=0$ the geometry is completely regular at the level of the curvature invariants. The general expressions are somewhat cumbersome; however, their behavior near $r=0$ is given by
\begin{subequations}
	\begin{align}
		\lim_{r\to 0}R&=4(\Lambda+\kappa)+\mathcal{O}(r^2)\,,\\
		\lim_{r\to 0}R_{\mu\nu}R^{\mu\nu}&=4(\Lambda+\kappa)^2+\mathcal{O}(r^2)\,,\\
		\lim_{r\to 0}R_{\mu\nu\alpha\beta}R^{\mu\nu\alpha\beta}&=\frac{8}{3}(\Lambda+\kappa)^3+\mathcal{O}(r^2)\,.
	\end{align}
\end{subequations}
Furthermore, for $q>3/2$, near the origin behaves as
\begin{equation}
	f(r)\sim 1-\frac{(\Lambda+\kappa)r^2}{3}+\mathcal{O}(r^4)\,,
\end{equation}
 which exhibits a local maximum at $r=0$, with $f(0)=1$, thereby indicating the presence of a de Sitter core with an effective cosmological constant shifted by the gravitational coupling constant.

\begin{figure}[h!]
        \centering
        \includegraphics[width=\linewidth]{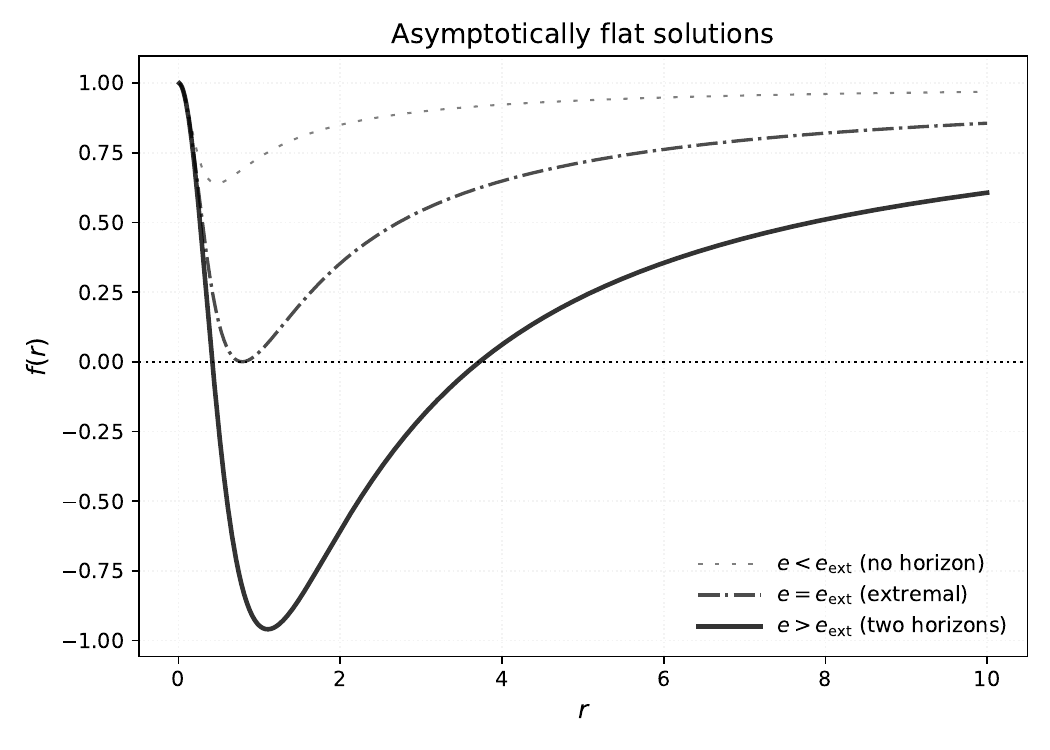}
        \caption{Gravitational potential $f(r)$ for asymptotically flat solutions. There is a precise value $e_{\text{ext}}$ where the black hole is extremal, in the sense that $f(r_{+})=f'(r_{+})=0$. Here, we have used $\kappa = 8\pi$, and $q=5/2$ for the plots.}
        \label{fig:horizon-flat}
    \end{figure}

\begin{figure}[h!]
        \centering
        \includegraphics[width=\linewidth]{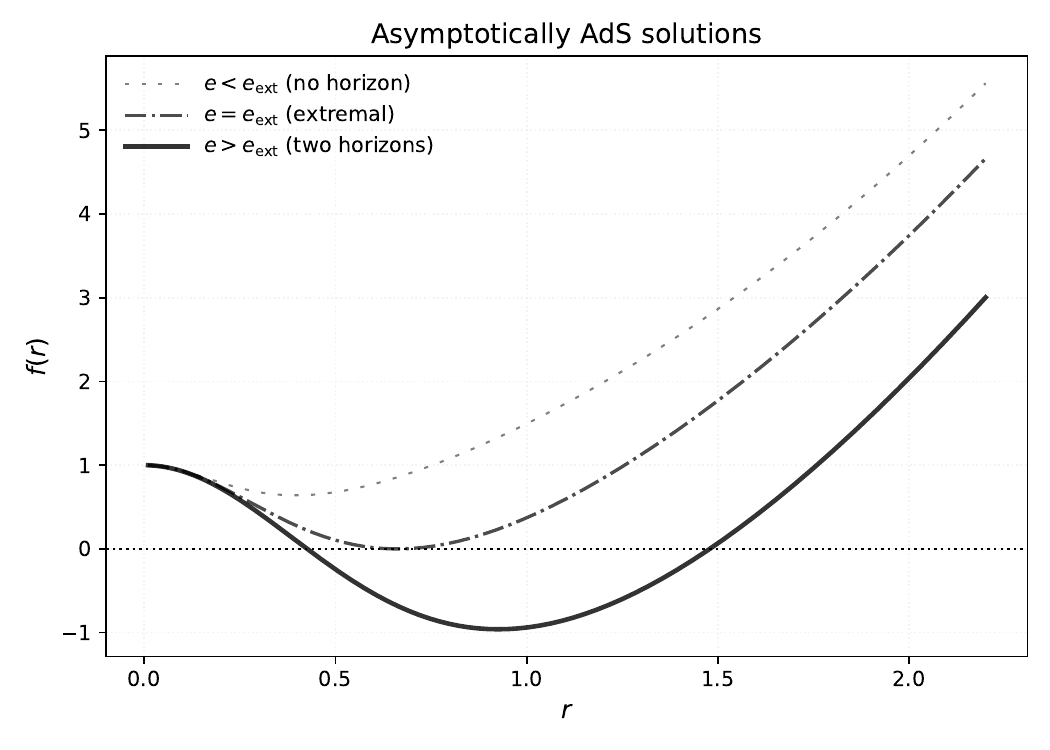}
        \caption{Gravitational potential $f(r)$ for asymptotically AdS solutions. We have used $\kappa=8\pi$, $q=5/2$, and $\Lambda=-3$ for the plots.}
        \label{fig:horizon-ads}
    \end{figure}

 For asymptotically flat (or asymptotically AdS) configurations, one has $f(r)\to 1$ (or $f(r)\to +\infty$) as $r\to\infty$, as can be seen from Eq.~\eqref{asympt-f-BH}. Hence, the existence of an event horizon depends on the underlying non-Abelian theory, through the coupling constant $e$. Indeed, since $f(r)$ decreases in a neighborhood of the origin, continuity together with the asymptotic behavior guarantees the existence of a local minimum. For sufficiently large values of $e$, this minimum becomes negative, so that $f(r)$ admits two real positive zeros, corresponding to the inner and outer horizons. In this regime, the non-Abelian theory supports a regular black hole geometry. This generic feature is presented in the following plots, Fig.~\ref{fig:horizon-flat} and Fig.~\ref{fig:horizon-ads}.


The simplest way to visualize the relation between $q$ and $e$ is through the extremality condition,
\begin{align}
    f(r_{+})= f'(r_{+})=0\,,
\end{align}
where $r=r_{+}$ is the black hole horizon. Take for instance the case $\Lambda=0$ and $q=5/2$, so that the spacetime is effectively the Bardeen metric,
\begin{align*}
    f(r) = 1 - \dfrac{\kappa r^2}{3}\left(\dfrac{e}{r^2+e}\right)^{\frac{3}{2}}\,.
\end{align*}
The metric function has a minimum at $r_{ext}=\sqrt{2e}$, where $f(r_{ext})= 1-\frac{2\sqrt{3}}{27}\kappa e$. Therefore, the analysis of the horizon structure is simplified to the following cases,
$$\begin{cases}
    \text{no horizon} &\text{if}\quad  e< \frac{27}{2\sqrt{3}\kappa}\,, \\
    \text{extremal BH} &\text{if}\quad e= \frac{27}{2\sqrt{3}\kappa}\,, \\
    \text{two horizons} &\text{if}\quad e> \frac{27}{2\sqrt{3}\kappa}\,.
\end{cases}$$
For larger half-interger values of $q$, the condition $f'(r)=0$ leads to algebraic equations of increasing degree, making the analytic treatment progressively less tractable. The situation becomes more involved when other values of $q$ are considered. Evenmore, when $\Lambda \neq 0$ and $q=5/2$, the equation $f'(r)=0$ already translates into a fith-order algebraic equation. Thus, we employ numerical methods to integrate the curve $e=e(q)$, as shown in Figure \ref{fig:extremal} below.
    \begin{figure}[h!]
        \centering
        \includegraphics[width=\linewidth]{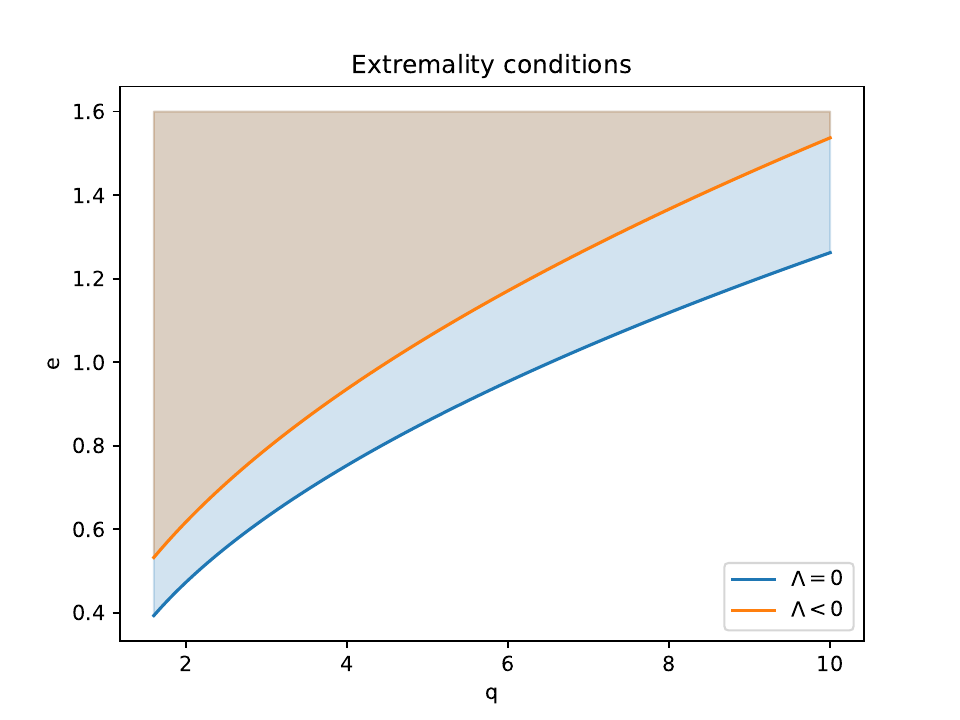}
        \caption{The solid curves describe the pair $(q,e)$ such that the regular black hole is extremal. The blue (orange) curve corresponds to the asymptotically flat (AdS) solution. In both cases, the upper-half plane corresponds to the pairs $(q,e)$ where the metric possesses both inner and outer horizons.}
        \label{fig:extremal}
    \end{figure}

It is worth emphasizing that, although regular black hole solutions have been constructed within Einstein--Yang--Mills theory, such configurations typically belong to effectively Abelian sectors of the theory. For instance, in the seminal work of Bartnik and McKinnon~\cite{Bartnik:1988am}, it was shown that the gauge field reduces to a $U(1)$ configuration, closely related to the Dirac monopole, thereby giving rise to an effective Reissner--Nordstr\"om geometry. By contrast, the solution presented here is genuinely non-Abelian in nature, since the gauge field does not reduce to an Abelian subgroup, and the regularization of the geometry emerges intrinsically from the non-Abelian sector, rather than from an effective Abelian truncation. Due to its non-Abelian nature, the meron circumvent the hypotheses from the theorem established in Ref.~\cite{Bronnikov:2000vy}. To the best of our knowledge, this represents the first analytical regular black hole solution supported by genuinely non-Abelian gauge fields.

\subsection{Meronic AdS wormholes instantons}
Another class of configurations that can be supported by meronic fields is that of Euclidean wormholes~\cite{Giddings:1987cg,Hawking:1987mz,Lavrelashvili:1987jg,Hosoya:1989zn,Marolf:2021kjc}. These solutions correspond to geometries with more than one disconnected boundary, which are connected through the bulk by a throat.

When Einstein gravity with a negative cosmological constant is coupled to Maxwell fields, the system does not support such configurations if the boundary has positive sectional curvature~\cite{Witten:1999xp,Cai:1999dqz}, as this leads to instabilities in the dual conformal field theory. As a result, a variety of Euclidean wormhole solutions have been constructed either by allowing for boundaries with negative curvature or by introducing appropriate matter content~\cite{Maldacena:2004rf,Buchel:2004rr,Arkani-Hamed:2007cpn}. Interestingly, this obstruction can be circumvented in the presence of non-Abelian fields. In particular, meronic configurations arising in Einstein--Yang--Mills system have been shown to support Euclidean wormholes~\cite{Maldacena:2004rf}. Motivated by this, it is natural to ask whether more general non-Abelian theories beyond Yang--Mills also admit such configurations.

To construct this class of solutions, we adopt a metric ansatz that share the isometries of the round three-sphere, namely
\begin{equation}\label{ds-wormhole}
	\dd s^2=\frac{\dd r^2}{f(r)}+(r^2+r_0^2)\dd\Omega^2_{(3)}\,,
\end{equation}
where $r_0\in\mathbb{R}\setminus\{0\}$ characterizes the radius of the wormhole throat, and
\begin{equation}
\dd\Omega^2_{(3)}=\frac{1}{4}\left(\sigma_1^2+\sigma_2^2+\sigma_3^2\right)\,,
\end{equation}
is the metric on the round three-sphere. For the gauge connection, we employ the meron-like ansatz given in Eq.~\eqref{A-theorem}, which automatically solves the field equations~\eqref{eom-A}, as a consquence of the Theorem \ref{theorem}. In this case, the invariant $Y$ vanishes identically, while
\begin{equation}
	X=-\frac{3}{2(r^2+r_0^2)^2}\,,
\end{equation}
which remains regular throughout the entire radial domain due to the presence of the wormhole throat. The Einstein equations~\eqref{Einstein-eqs}, are solved by
\begin{equation}\label{fsol-WH}
	f(r)=\left(1+\frac{r_0^2}{r^2}\right)\left(1+\frac{r^2+r_0^2}{\ell^2}+\frac{\kappa(r^2+r_0^2)}{3}\Lag\right)\,.
\end{equation}

To ensure that the geometry is regular at the wormhole throat, it is instructive to examine the behavior of the metric function near $r=0$ (see, e.g., Ref.~\cite{Marolf:2021kjc}). Expanding Eq.~\eqref{fsol-WH} around this point yields
\begin{equation}
\begin{split}
	f(r)&\sim \frac{r_0^2[\kappa\ell^2r_0^2\Lag_0+3(\ell^2+r_0^2)]}{3\ell^2 r^2}+\frac{\kappa r_0^4}{3r}\eval{\dv{\Lag}{r}}_{r=0}+\frac{\kappa r_0^4}{6}\eval{\dv[2]{\Lag}{r}}_{r=0}\\
	&~~+\frac{2\kappa\ell^2r_0^2\Lag_0+3(\ell^2+2r_0^2)}{3\ell^2}+\mathcal{O}(r)\,.
\end{split}
\end{equation}
It then follows immediately that regularity at the throat requires the coefficients of the divergent terms to vanish. Consequently, the throat radius is not a free parameter but is instead fixed in terms of the couplings of the theory according to 

\begin{equation}\label{regularity-cond-WH}
	r_0^2=-\frac{3\ell^2}{\left(3+\kappa\ell^2\Lag_0\right)}\,,\quad \text{with}\quad \Lag_0<-\frac{3}{\kappa\ell^2}\,,
\end{equation}
where we have defined $\Lag_0\equiv \Lag(r=0)$. In addition, one must require
\begin{equation}
	\eval{\dv{\Lag}{r}}_{r=0}=0\,.
\end{equation}
Under these boundary conditions, the metric function in Eq.~\eqref{fsol-WH} reduces to
\begin{equation}
	f(r)=\frac{(\kappa\ell^2\Lag+3)r^2}{3\ell^2}+\frac{\kappa\ell^2(\Lag_0-2\Lag)-3}{(\kappa\ell^2\Lag_0+3)}-\frac{3\kappa\ell^4(\Lag_0-\Lag)}{(\kappa\ell^2\Lag_0+3)^2r^2}\,.	
\end{equation}
In general, this metric function is positive definite only within a certain region of parameter space, which depends on the specific non-Abelian theory under consideration. In such a case, this corresponds to a completely regular solution, both at the level of the gauge invariants and the curvature invariants, provided that the throat radius $r_0$ is nonvanishing.

We have explicitly shown that the non-Abelian extensions of ModMax~\cite{Cirilo-Lombardo:2023poc,Canfora:2025gwm}, Euler--Heisenberg~\cite{Dowker:1975tf,Duff:1975ue,Dunne:2004nc}, Born--Infeld~\cite{Tseytlin:1997csa}, and the ABG theories satisfy these conditions, thereby demonstrating that such configurations can also be realized in gauge theories beyond Yang--Mills.

\subsubsection{Non-Abelian ModMax meronic wormholes}

As a representative example, let us consider the non-Abelian extension of the ModMax nonlinear electrodynamics~\cite{Bandos:2020jsw,Kosyakov:2020wxv,Ayon-Beato:2024vph} introduced in Ref.~\cite{Cirilo-Lombardo:2023poc} (see also~\cite{Canfora:2025gwm}), coupled to Einstein gravity with a negative cosmological constant, $\Lambda=-3/\ell^2$, where $\ell$ denotes the AdS radius. The corresponding action principle is given by
\begin{equation}
\label{eq:wh-action}
	I[g,A]=\int_\mathcal{M}\dd^4x\sqrt{|g|}\left[\frac{R-2\Lambda}{2\kappa}+X\cosh{\gamma}-\sqrt{X^2-Y^2}\sinh(\gamma)\right]+\frac{1}{\kappa}\int_{\partial\mathcal{M}}\dd^3x\sqrt{|h|}\mathcal{K}+I_{\rm ct}\,.
\end{equation}
Here, the first boundary contribution corresponds to the Gibbons-Hawking-York (GHY) term~\cite{Gibbons:1976ue,York:1972sj}, which is required to ensure a well-posed variational principle under Dirichlet boundary conditions for the induced metric $h_{a b}$. It can be derived by introducing Gaussian-normal coordinates, in which the bulk metric takes the form
\begin{equation}
	\dd s^2=N^2(r)\dd r^2+h_{ab}(r,x)\dd x^{a}\dd x^b\,,
\end{equation}
where $\{x^{a}\}$ denote the boundary coordinates, $N(r)$ is the lapse function, and $h_{ab}$ is the induced metric on the codimension-one hypersurfaces of constant $r$. Throughout this section, Latin indices from the beginning of the alphabet refer exclusively to boundary quantities. With this foliation, the extrinsic curvature is determined by the radial derivative of the induced metric, $\mathcal{K}_{a b}=-\frac{1}{2N}\partial_r h_{a b}$, whose trace is given by $\mathcal{K}=\mathcal{K}_{a b}h^{a b}$. The second boundary contribution, $I_{\rm ct}$, consists of the counterterms required to render both the on-shell action and its variation finite for asymptotically AdS spacetimes. It is given by
\begin{equation}
	I_{\rm ct}=\frac{1}{\kappa}\int_{\partial\mathcal{M}}\dd^3x\sqrt{|h|}\left(z_0+z_1\mathcal{R}\right)\,,
\end{equation}
where $\mathcal{R}(h)$ is the intrinsic curvature scalar associated with the induced metric $h_{ab}$; the latter is used to raise and lower the boundary indices. The coefficients $z_0$ and $z_1$ are fixed by requiring the total action to be finite. 

 Adopting the wormhole ansatz~\eqref{ds-wormhole}, the Einstein equations are solved by Eq.~\eqref{fsol-WH}, which in the present case reduces to\begin{equation}
	f(r)=\frac{(r^2+r_0^2)(\ell^2+r^2+r_0^2)}{\ell^2r^2}-\frac{\kappa e^\gamma}{2r^2}\,.
\end{equation}
To ensure regularity at the wormhole throat, we first evaluate the on-shell matter Lagrangian at $r=0$, obtaining
\begin{equation}
	\Lag_0=-\frac{3}{2}\frac{e^\gamma}{r_0^4}\,.
\end{equation}
Substituting this expression into the regularity condition~\eqref{regularity-cond-WH} determines the throat radius as
\begin{equation}
	r_0^2=\frac{\ell}{2}\left(\sqrt{\ell^2+2\kappa e^\gamma}-\ell\right)\,.
\end{equation}
Note that the throat of the wormhole depend on the ModMax parameter $\gamma$, as anticipated in Eq.~\eqref{regularity-cond-WH}. Consequently, the metric function simplifies to the remarkably compact form
\begin{equation}
	f(r)=\frac{r^2}{\ell^2}+\frac{\sqrt{\ell^2+2\kappa e^\gamma}}{\ell}\,,
\end{equation}
which is positive definite for all values of the parameters. This geometry is free of curvature and gauge singularities, as all corresponding invariants remain finite provided that $r_0\neq 0$.

We turn now to the calculation of the Euclidean action. To this end, we first note that both the bulk action density and the counterterm integrand are even functions of the radial coordinate $r$. Consequently, their contributions are simply twice the corresponding integrals evaluated over the interval $r\in[0,\infty)$. This is not the case for the GHY boundary term, whose contribution must instead be evaluated explicitly on each of the two asymptotic boundaries. Taking into account that one has to choose the outward normal vector at each boundary, the action is finite by choosing
\begin{align}
    z_0 = -\dfrac{2}{\ell}\,,\quad z_1 = -\dfrac{\ell}{2}\,,
\end{align}
which coincides with the standard values for AdS$_4$~\cite{Balasubramanian:1999re,Emparan:1999pm}. Using these values, the action \eqref{eq:wh-action} reads
\begin{align}
    I[g,A] = \dfrac{4\pi^2 \sqrt{\ell  s}}{\kappa} \left[(\ell-s)K(m) -2\ell E(m) \right]\,,
\end{align}
where $s=\sqrt{\ell^2+2\kappa e^{\gamma}}$, $m= \dfrac{s+\ell}{2s}$, and $K(m), E(m)$ are the
complete elliptic integrals of the first and second kind, 
\begin{align}
    K(m) = \int_0^{\pi/2}\dfrac{\dd\theta}{\sqrt{1-m\sin^2\theta}}\,,\qquad
    E(m) = \int_0^{\pi/2}\sqrt{1-m\sin^2\theta}\;\dd\theta\,,
\end{align}
respectively.

\section{Discussion}\label{sec:Discussion}
Universal configurations are scarce and of considerable physical interest, as they are, in a certain sense, quantum protected, being insensitive to quantum corrections, among other deformations. In this work, we have shown that meronic configurations, originally discovered in flat space within Yang--Mills theory~\cite{deAlfaro:1976qet}, can be extended to a broader class of non-Abelian gauge theories. This configurations arise naturally in parity invariat theories, thereby revealing the existence of a universal sector in non-Abelian gauge dynamics. We have further demonstrated that they can be consistently defined on constant--curvature spaces, not necessarily restricted to flat backgrounds. As an explicit example, we consider a family of non-Abelian theories inspired by the ABG nonlinear electrodynamics~\cite{Ayon-Beato:2000mjt}, for which the meronic configurations acquire a finite on-shell action despite being singular at the origin, in contrast to standard Yang--Mills theory.

When gravitational backreaction is taken into account, this universal character is preserved for both black hole and Euclidean wormhole solutions dressed by merons, and we have shown that they can be extended beyond the Einstein--Yang--Mills theory. In this setting, the singular behavior typically exhibited by merons is resolved by the global structure of spacetime: it is hidden behind an event horizon in the case of black holes, or smoothed out by the presence of a throat in the case of wormholes. In the latter case, this leads to configurations that are completely regular at the level of both curvature and gauge invariants. However, in the case of black holes, since we have shown that they are supported by a broad class of non-Abelian theories, we focus in particular on the non-Abelian extension of the ABG theory. In this framework, we show that the theory admits regular charged black hole solutions supported by genuinely non--Abelian fields.

In Section \ref{sec:Self-gravitating merons}, we have shown that the meron gravitates and backreacts, but it is unresponsive to the nonlinearity. This is manifestly different from the abelian case, where for nonlinear theories with electric fields the gauge potential departs from the Coulomb behavior of the Reissner-Norstr\"om black hole due to its nonlinear nature.

These results provide strong evidence that merons capture robust nonperturbative features of non-Abelian gauge theories in a model-independent way. From this perspective, they provide access to sectors of the theory that remain stable under deformations and higher--order corrections. This further supports the view that such configurations are effectively quantum protected, and therefore of particular relevance in the study of non-Abelian effective field theories.

Several directions naturally emerge from this work. First, it would be interesting to extend our analysis to more general gauge groups beyond $SU(2)$. In particular, exploring the case of $SU(N)$ could reveal a richer structure, including new classes of meronic configurations and a broader spectrum of topological sectors. Second, it would be worthwhile to investigate the realization of the spin from isospin effect, which has been previously studied in the context of meronic black holes in Einstein--Yang--Mills theory~\cite{Canfora:2012ap}. It would be especially interesting to understand how this mechanism manifests in more general frameworks, such as the theory defined in Eq.~\eqref{non-Abelian ABG}, particularly in the presence of regular black hole solutions, and even in the case of larger gauge groups, where it has been shown to have a direct impact on this phenomenon~\cite{Canfora:2022nso,Aviles:2026fsf}. This could have important implications within the context of the AdS/CFT correspondence, as it opens the possibility of realizing fermionic excitations on the boundary without introducing fundamental fermionic degrees of freedom in the bulk. It would also be worthwhile to investigate whether meronic configurations remain solutions of gauge theories constructed from other independent gauge invariants of the $SU(2)$ gauge group, beyond those considered in Eq.~\eqref{X-Y}. One can invoke some underlying symmetry arguments to anticipate a partial answer. For example, one can check that the field strength $F$ \eqref{F} is invariant under the map $\lambda \mapsto 1-\lambda$. The unique fixed point of such transformation is precisely the meron. Therefore, one may expect this meronic configuration to remain a critical point of the equations of motion \eqref{eom-A} for theories built from more general invariants of the field strength. From what we have learned from Section \ref{sec:merons on constant curvature spaces}, we conjecture that parity-even actions stay stationary at $\lambda=1/2$, regardless of considering covariant derivatives, higher-order traces, or different contractions. A positive general answer would considerably broaden the scope of the present results, suggesting that merons may define a genuinely universal sector of gauge theories. Finally, it would be natural to explore the existence of Lorentzian counterparts of the Euclidean wormholes discussed in this work. In particular, one may ask whether traversable wormholes can be supported by genuinely non-Abelian matter fields within theories beyond the Einstein--Yang--Mills framework. We leave these directions for future work.

\begin{acknowledgments}
We thank Fabrizio Canfora, Crist\'obal Corral, and Julio Oliva for helpful discussions and support. We also thank Eloy Ay\'on-Beato and Pedro A. S\'anchez for enlightening conversations. The authors are sincerely grateful to the anonymous referee for the careful reading of the manuscript and for the many insightful comments and suggestions, which have substantially improved the quality and clarity of this work. The work of L.G. is partially supported by the Agencia Nacional de Investigaci\'on y Desarrollo (ANID) through Fondecyt Iniciaci\'on grant No.11260910.
\end{acknowledgments}

\bibliographystyle{JHEP}
\bibliography{biblio.bib}

\end{document}